\documentstyle[12pt,world_sci]{article}
\def\eslt{E\llap/_T}
\def\to{\rightarrow}
\def\Re{{\cal R \mskip-4mu \lower.1ex \hbox{\it e}}\,}
\def\Im{{\cal I \mskip-5mu \lower.1ex \hbox{\it m}}\,}

\def\te{\tilde e}
\def\alt{\stackrel{<}{\sim}}
\def\agt{\stackrel{>}{\sim}}
\def\tl{\tilde l}

\def\tg{\tilde g}

\def\tnu{\tilde \nu}
\def\tell{\tilde \ell}
\def\tq{\tilde q}
\def\tb{\tilde b}
\def\tt{\tilde t}
\def\tw{\widetilde W}
\def\tz{\widetilde Z}

\begin{document}
\begin{flushright}
UH-511-820-95
\end{flushright}
\title{{\bf RECENT DEVELOPMENTS IN SUPERSYMMETRY SEARCH
STRATEGIES}\thanks{Talk presented at the Eleventh Department of Atomic Energy
Symposium, Visvabharati University, Santiniketan, Dec. 28, 1994 - Jan. 2,
1995.}}
\author{XERXES TATA\\
{\em Department of Physics and Astronomy, University of Hawaii,\\
Honolulu, HI 96822, USA}}
%
\maketitle
\setlength{\baselineskip}{2.6ex}

\begin{center}
\parbox{13.0cm}
{\begin{center} ABSTRACT \end{center}
{\small \hspace*{0.3cm} After a quick review of the framework for
phenomenological analyses of supersymmetry, we
summarize
current limits on
supersymmetric particle masses and discuss strategies for their
searches at the Fermilab Tevatron and the LHC. We also discuss the sense
in which such searches as well as those that may be carried out at LEP
complement one another. Finally, we touch upon the prospects for
more ambitious measurements
such as those of sparticle masses and couplings. Such measurements could
(a) help pin down model
parameters, (b) perhaps, serve to test our ideas of physics at very high
energy, and (c) provide the most direct test of supersymmetry.  }}
\end{center}

\section{Introduction and Model Framework}

Aesthetic issues aside, the
interest in low energy SUSY phenomenology
was originally driven by the
recognition that SUSY can stablize the gauge hierarchy provided sparticles
are lighter than $O$(1 TeV). More recently,
the realization that the precise measurements of the
gauge couplings at LEP are in agreement with the minimal supersymmetric
model but {\it incompatible}\cite{UNIF} with its minimal non-supersymmetric
counterpart
has led several groups\cite{SPECTRA} to re-examine the expectations for
sparticle
masses and mixing patterns within the theoretically appealing supergravity
(SUGRA) framework.
Experimental constraints from flavour changing neutral currents can
readily be accommodated\cite{FCNC} and, assuming $R$-parity is unbroken,
SUSY models include a viable candidate
for dark matter\cite{DM}.

Most phenomenological analyses have been performed within the framework
of the Minimal Supersymmetric Model (MSSM),
which is a supersymmetrized version of the SM with the
least number of additional new particles and interactions necessary
for phenomenology\cite{REV}. It is assumed that $R$-parity is exactly conserved
so that the lightest SUSY particle (LSP) is stable. If it is further
assumed to be the dark matter in our galactic halo, the LSP can only be the
lightest neutralino.
Supersymmetry breaking is parametrized by the introduction
of soft SUSY breaking mass terms and interactions consistent with
the SM gauge group. A general analysis requires an unmanageably large number
of parameters and, without further assumptions, the phenomenology is
intractable. This was not a problem in the earliest analyses when it
was assumed that a sparticle always directly decays to ordinary particle(s)
and the LSP; in this simple case, the rates for SUSY signals were essentially
fixed by the production process, {\it i.e.} by the mass (and, possibly mixing
angles) of the sparticle being searched for. It was subsequently recognized
that heavy sparticles decay via a complicated cascade\cite{CAS} that terminates
in
the LSP so that the resulting signals also depend on the masses and
couplings of all sparticles that participate in the decay cascade.

 Motivated by grand unified theories
(GUTs), most early collider studies of SUSY\cite{EARLY,BTW,SDC,GEM}
(and also those of today)
assumed that the three running gaugino masses originate in a
common gaugino mass at some high scale, so that the ratios of these masses
evaluated at the weak scale, are proportional to the corresponding
ratios of the (known) squared gauge couplings; the $SU(2)$ and $U(1)$ gaugino
masses are then determined in terms of $m_{\tg}$. To further
reduce the number of parameters,
it was usually assumed that the sfermions were all degenerate
(this was motivated by supergravity (SUGRA) model studies); the Higgs
boson sector (at tree level) is completely determined by the mass ($m_{H_p}$)
of the neutral pseudoscalar along with
the parameters $\tan\beta$,
the ratio of the two vacuum expectation values and $\mu$, the
superpotential Higgsino mass which also enter the gaugino-Higgsino sector of
the model. Together with the
$A$-parameter(s) which affects mainly the
phenomenology
of third generation squarks, this parameter set
completely specifies the model. We will refer to this framework
for sparticle masses and mixings as the MSSM.

In view of the fact that the simplifying assumptions underlying these MSSM
analyses are motivated by supergravity (SUGRA)
GUT models, it seems reasonable to
seriously explore their implications. Several phenomenological
analyses of SUGRA models have recently appeared in the
literature\cite{MUR,LOPEZ,BDKNT,BCMPT}.
These models are based on assumptions about the symmetries of interactions
at an ultra-high energy scale where the physics is simple. We will not
discuss these here, but instead refer the reader to the literature.
The effective Lagrangian of the low energy theory
is completely specified in terms
of just four parameters, renormalized at the high scale, where the assumed
symmetries relate many masses and couplings. The evolution of these parameters
down to the weak scale relevant to phenomenology\footnote{This is essential
to sum large the logarithms arising from the disparity of the two scales.}
then leads to a diverse pattern of sparticle masses and mixings. Typically,
one finds that

\begin{itemize}
\item the first two generations of squarks are
(approximately)
degenerate so that these models automatically satisfy constraints\cite{FCNC}
from the
non-observation of flavour changing neutral currents in the kaon sector, while
$\tt_1$, the lighter $t$-squark and
$\tb_1 \sim\tb_L$, the lighter $b$-squark may be substantially lighter. This
can lead to distinctive third generation phenomenology\cite{GUN,BST},
as well as lead to
enhanced production of third generation (s)particles, both directly and via
gluino decays.

\item Sleptons are lighter than the
squarks, and may be much lighter if gluinos and
squarks (of the first two generations) have comparable masses. This can lead
to an enhancement of leptonic decays of neutralinos, and sometimes, also
charginos which, in turn, leads to increased cross sections for
leptonic signals via which SUSY might reveal itself at colliders.

\item An especially attractive property of this framework is that the evolution
of the model parameters from the high scale down to the weak scale also
triggers the breakdown of electroweak symmetry. Some caution must
be exercised in the choice of input parameters to ensure that colour
and electromagnetic gauge invariances are not spontaneously broken. This
cannot, therefore, be regarded\footnote{It should also be pointed out
that this requires that the input parameters at the ultra-high scale be
O(1~TeV). A complete understanding should include the explanation of why
the values of these parameters renormalized at a scale $\sim 10^{16-18}$~GeV
should be just $\sim 1$~TeV.} as an explanation of the manifest
QCD and electromagnetic gauge
invariances. Nonetheless, the fact that the correct breakdown of
SU(2)$\times$U(1) can be obtained over wide ranges of parameters must be
regarded as a success of the model.
The magnitude of $\mu$ is fixed to reproduce the measured $Z$ boson mass.

\end{itemize}

The particle/sparticle masses and couplings are determined by a few
parameters (along with $m_t$) which may be chosen to be:
\begin{eqnarray}
m_{0},\ m_{1/2},\ A_0,\ \tan\beta\ {\rm and}\ sgn(\mu ).
\end{eqnarray}

The recent version\cite{ISAJET} of the supersymmetry simulation program ISAJET
7.13
now gives the reader the option to either
input the supergravity parameter set (1)
above, or to use the more extended MSSM parameter set for the analysis.
The former choice automatically incorporates sparticle masses and decay
patterns as given by the restricted SUGRA framework. Because the model
is specified by only four input parameters, various experimental observables
become correlated. As we will see,
these correlations provide non-trivial tests of the
underlying extrapolation of our knowledge of physics to very high scales
(which could ultimately prove to be incorrect). The sensitivity of the model
predictions to SUGRA assumptions can be tested by using the optional
MSSM inputs. In this way ISAJET can also be used to simulate models with
non-universal soft-breaking sfermion terms.

\section{Current Status of Supersymmetry}

There is no direct evidence for sparticles in high energy collisions. This
is not (yet) a cause for despair since,
if we regard fine-tuning issue as the
motivation for low energy SUSY, we would
typically expect sparticle masses around 100-1000~GeV. Since SUSY theories
are of the decoupling type in that virtual effects of sparticles decouple
as $m_{SUSY}\rightarrow\infty$, the
non-observation\cite{NODEV}
of any significant
deviations from the Standard Model (SM) in precision experiments at
LEP is compatible with (though not an argument for) low energy SUSY.

The most
straightforward limits on masses of sparticles with gauge couplings to the $Z$
come from a measurement of its total width; these limits on the mass of
the charged slepton and sneutrino,
the squark and the chargino ($\tw_1$), which are
independent of how sparticles decay, are a few GeV below
the kinematic bound $M_Z/2$. Exclusive searches and the measurement of
the invisible and leptonic widths of the $Z$ yield (sparticle
decay-dependent) lower limits\cite{LEPREV}
even closer to $M_Z/2$. Since
neutralinos ($\tz_i)$
couple to the $Z$ only via their Higgsino components, the bounds on their
masses are sensitive to model parameters which determine the mixing patterns.

The search
for squarks and gluinos, because of their strong interactions, is best
performed
at hadron colliders. The non-observation of an excess of $\eslt$ events
at the Tevatron has enabled the D0 collaboration\cite{DZERO} to infer lower
limits of about 150~GeV on their masses, improving on the earlier limit of
$\sim 100$~GeV obtained by CDF\cite{CDF}; if $m_{\tq}=m_{\tg}$, the mass bound
improves
to 205~GeV. These bounds, which assume ten flavours of degenerate squarks,
have some sensitivity to the model parameters
which determine the cascade decay patterns of the $\tq$ and $\tg$. Within the
minimal model, the Tevatron and LEP bounds together imply\cite{ROSZ}
a lower
bound just above 20~GeV on the mass of the $\tz_1$ which, because $\tz_1$
is assumed to be the stable LSP, may have some cosmological
significance.

There are also constraints from ``low energy'' experiments and from cosmology.
Some judgement must be exercised in evaluating these constraints which can
frequently be evaded by minor modifications of the model. For example, the
constraint from an over-abundance of LSPs leading to too short-lived a
universe might be evaded by allowing a tiny violation of R-parity
conservation without impact on collider searches.\footnote{It is perfectly
possible to construct phenomenologically viable models that violate
$R$-parity conservation. We do not discuss these here for reasons of time.}
Proton decay
constraints\cite{PDK}
are sensitive to details about GUT scale physics. Finally, we remark that
flavour violating decays, {\it e.g.} inclusive $b \to s\gamma$ recently
measured by CLEO\cite{CLEO} serve to restrict the {\it sum} of new physics
contributions. Since it is frequently possible for partial cancellations
amongst these, care must be exercised in extracting limits on sparticle
masses from these data.

\section{SUSY Search in the 1990's}

The LEP collider is expected to enter its second phase of operation
around the end of next year when its energy will be upgraded to 175-200~GeV.
Given a data sample of $O$(100 $pb^{-1}$), the
clean environment of $e^+e^-$ collisions will readily enable
experimentalists to search for charginos, sleptons, squarks\cite{DION} and even
Higgs bosons\cite{WU} with masses up to 85-90~GeV ($b$-tagging capability may
be necessary if $m_H \simeq M_Z$). The corresponding
mass reach for neutralinos
is sensitive to (model-dependent) mixing angles. The process $e^+e^- \to
\tz_1\tz_2$ may enable experiments to probe parameter values that cannot
be probed via chargino pair production.

At the Tevatron, gluino and squark production processes dominate
the production of the less massive charginos and neutralinos for
gluino masses up to 320~GeV (200~GeV) if $m_{\tq} = m_{\tg}$ ($m_{\tq} =
2m_{\tg}$). Thus, while gluinos and squark signals may be the best way
to search for SUSY at present, the ultimate reach of (a very high luminosity
upgrade of) the Tevatron may well be determined by the trilepton signal from
$\tw_1\tz_2$ production. The production of charginos or neutralinos
in association with squarks and gluinos is never dominant.

The D0 and CDF experiments have collectively accumulated an
integrated luminosity in excess of 100 $pb^{-1}$
and will continue to accumulate data for several more months.
In addition to extending the $\eslt$ search region, the
large increase in the data sample (relative to their earlier analyses
which were based on a $\sim 10$ $pb^{-1}$) should enable them to
({\it i}) perform gluino and squark searches
via multilepton events from their cascade
decays, ({\it ii}) search for $\tw_1\tz_2$ production via isolated
trilepton events free of jet activity, and ({\it iii}) search for the
lighter $t$-squark, $\tt_1$. The Tevatron is unlikely
to be able to detect sleptons beyond the range of LEP\cite{SLEP}.

{\it Multilepton Signals from Gluinos and Squarks.} The conventional $\eslt$
search for $\tg$ and $\tq$ is background limited. Even with an integrated
luminosity of 1 $fb^{-1}$ that should be available with the Main Injector
upgrade of the Tevatron, we anticipate a maximum reach of $\sim 270$~GeV
($\sim 350$~GeV) if $m_{\tq} >> m_{\tg}$ ($m_{\tq} \simeq m_{\tg}$) in this
channel\cite{KAMON,DPF}.
Heavy gluinos and squarks can also decay via the $\tw_1$ and
$\tz_2$ modes which, unless suppressed by phase space,
frequently dominate the decays of $\tq_L$ and $\tg$. The subsequent leptonic
decays of the $\tw_1$ and $\tz_2$ yield events with hard jets accompanied by
1-3 isolated, hard leptons and $\eslt$. The cross sections for various event
topologies, with cuts described in Ref.[31] is shown in Fig.~1. Our
simulation includes the simultaneous production of all sparticles. Thus, for
very large values of $m_{\tg}$, the cross sections are dominated by events
from chargino and neutralino production.
\begin{figure}[hbt]
\vspace{90mm}
\caption{Tevatron ($\protect{\sqrt{s}}=1.8$~TeV) cross sections
in $fb$ for various event topologies after the cuts
of Ref.[31] from which this figure is taken. We take $\mu=-m_{\tg}$,
$\tan\beta=2$, $m_{H_p}= 500$~GeV, and set
the weak scale $A$ parameters equal to $-m_{\tq}$. The $\eslt$ cross sections
are
shown by diamonds, 1$\ell$ cross sections by crosses, $\ell^+\ell^-$
cross sections by x's and the same sign dilepton cross sections by squares.
The dotted and dashed curves show the trilepton and 4$\ell$ cross sections,
respectively.
The $m_{\tg}=150$~GeV case in Fig.~1{\it b} is excluded by LEP constraints
since this scenario has $m_{\tnu}=26$~GeV.}
\end{figure}

While there are substantial
backgrounds to $1\ell$ and $\ell^+\ell^-$ event topologies, the {\it physics}
backgrounds in the $\ell^{\pm}\ell^{\pm}$ and $3\ell$ channels are essentially
negligible, so that these search channels are essentially rate limited. These
channels, which at the Main Injector have a reach\cite{BKTGL} of 230-400~GeV
(for $\mu < 0$)
depending
on $m_{\tq}$ and $m_{\tg}$, provide complementary ways of searching for
gluinos and squarks at the Tevatron, and because they are free of SM
backgrounds, may even prove superior if gluinos and squarks are very heavy,
and the branching fraction for the decay $\tz_2 \to \ell^+\ell^- \tz_1$ is
not strongly suppressed.\footnote{Due to complicated interference effects,
a strong suppression of this decay occurs for
certain ranges of parameters if $\mu > 0$, so that the reach in especially
the 3$\ell$ channel requires further study.}

{\it Search for Isolated Trilepton Events.} Associated $\tw_1\tz_2$ production
which occurs by $s$-channel $W^*$ and $t$-channel $\tq$ exchanges, followed by
the leptonic decays of $\tw_1$ and $\tz_2$ results in isolated trilepton plus
$\eslt$ events, with hadronic activity only from QCD radiation. SM
backgrounds to the $3\ell + n_{jet}\leq 1$ signal are negligible,
assuming that $WZ$ events can be vetoed with
high efficiency by requiring $m_{\ell\bar{\ell}}\not=M_Z$ within experimental
resolution; a conclusive observation of a handful of such events would,
therefore,
be a signal for new physics. The branching fraction for leptonic decays
of $\tz_2$, and sometimes also of $\tw_1$, and hence, this signal,
is enhanced, especially for negative values of $\mu$,
if $m_{\tell} << m_{\tq}$ and the $\tz_2\tz_1Z$ coupling
is dynamically
suppressed; this situation frequently occurs in SUGRA type models, if
$m_{\tq} \simeq m_{\tg}$. On the other hand, if $\mu > 0$ this decay
may be strongly suppressed.
Preliminary analyses by the CDF and D0
experiments\cite{WINO}
(for large values of the Higgsino mass parameter, $\mu$)
are already competitive with bounds from LEP.
These experiments may soon explore\cite{BKTWIN,KAMON}
parameter ranges not accessible at LEP
and, under favourable circumstances, may be competitive with LEP II during the
Main Injector Era. The reach of
the Main Injector and of TeV$^*$, the proposed high luminosity upgrade of
the Tevatron designed to provide an integrated luminosity of
$\sim$ 25~$fb^{-1}$, in the $3\ell$ channel has recently been
studied\cite{KAMON,TEVSTAR,DPF}
within the SUGRA framework.
\begin{figure}[htb]
\vspace{170mm}
\caption{Regions of the $m_0-m_{1/2}$ plane where the trilepton signals
are observable above background at the Tevatron Main Injector upgrade
(1~$fb^{-1})$ and at the proposed TeV$^*$ upgrade with an integrated
luminosity of 25~$fb^{-1}$. This figure has been produced by C-H.~Chen.}
\end{figure}

The region of the $m_0-m_{1/2}$ plane where the signals may be statistically
significant is shown in
Fig.~2 for $\tan\beta=2$ and $A_0=0$. The solid squares denote where the
signal should be observable at the 5$\sigma$ level with an integrated
luminosity of 1 $fb^{-1}$. The remaining squares show the
corresponding regions where the signal is expected to be
observable at the TeV$^*$ upgrade. We see from Fig.~2{\it a} that for
$\mu < 0$,
the clean trilepton signal should be observable at
the 5$\sigma$ level for  most values of parameters where the ``spoiler'' decays
$\tz_2 \to \tz_1 H_{\ell}$ are not kinematically allowed. In contrast, for
positive values of $\mu$ shown in Fig.~2{\it b},
there is a large region for 400~GeV$\stackrel{<}{\sim} m_0 \stackrel{<}{\sim}$
1000~GeV which
even the TeV$^*$ may not be able to probe even for charginos just
beyond the current LEP reach\footnote{Our conclusion is at variance with what
has been stated in the literature\cite{KAMON}, and is because of the strong
suppression of the neutralino branching fraction mentioned above.}.

{\it Searching for Top Squarks at the Tevatron.}
The large Yukawa interactions of the top family which mix $\tt_L$ and $\tt_R$
serve to reduce the mass of $\tt_1$, the lighter of the
two mass eigenstates. In fact, it is theoretically possible that
$\tt_1$ is essentially massless with other squarks and gluinos all too heavy
to be produced at the Tevatron.
The Tevatron lower limits on $m_{\tq}$, are derived assuming ten
degenerate squark flavours, and so are not applicable to $\tt_1$. Currently,
the best limit, $m_{\tt_1}\agt M_Z/2$, comes from LEP experiments; this bound
can be evaded if the stop mixing angle and $m_{\tt_1}-m_{\tz_1}$ are both
fine-tuned, a possibility we do not entertain here.

The signals from top squark production depend on its decay patterns. For
$m_{\tt_1}< 125$~GeV, the range of interest at the Tevatron,
$\tt_1$ decays via the loop-mediated mode, $\tt_1\to c\tz_1$
when the
tree level decay $\tt_1\to b\tw_1$ is kinematically forbidden\cite{HK}.
Stop pair production is then signalled by $\eslt$ events from its direct
decays to the LSP.
With a data sample of 100 $pb^{-1}$,
Tevatron experiments should be able\cite{BST}
to probe stop masses up to 80-100~GeV,
significantly beyond the present bounds.
The tree level chargino
mode  dominates stop decays whenever it is kinematically allowed.
The subsequent leptonic
decay of one (or both) of the charginos lead to single lepton (dilepton)
+ $b$-jet(s)+$\eslt$ events, very similar to those expected from $t\bar{t}$
pair production. Top production is thus a formidable
background to the stop signal\cite{GUN}.
Special cuts need to be devised to separate stop
from top events.
Since
stops accessible at the Tevatron are considerably lighter than
$m_t$, and
because the chargino, unlike $W$, decays via three body modes into
a {\it massive} LSP, stop events are generally softer than top events. It has
been shown\cite{BST} that by requiring $m_T$($\ell\eslt$)$<45$~GeV
($p_T(\ell^+)+p_T(\ell^-)+\eslt<100$~GeV), in addition to other canonical
cuts, stops with masses up to about 100~GeV should be detectable
in the $1\ell$ (dilepton) channel by the
end of the current Tevatron run;
for top squark detection in the single lepton channel, sufficient
$b$-tagging capability is also required.

\section{Supersymmetry Searches at Supercolliders}

We have seen above that
direct searches at the Main Injector and LEP II will probe
sparticle masses between 80-300~GeV; even assuming the gaugino mass
unification condition, the
chargino search, by inference, will
probe gluino masses at most up to about 400-500~GeV.
Indeed while the TeV$^*$
may be able to probe sparticle masses significantly beyond the reach of
other collider facilities of the era for favourable values of parameters,
it should be kept in mind that it may not be able to probe (even indirectly
via the clean trilepton channel) gluino masses much beyond about 250-300~GeV if
squarks are heavy and $\mu$ is positive.
Since the SUSY mass scale
could easily be as large as 1~TeV, it will, unless sparticles have already been
discovered, be up to
supercolliders such as the LHC at CERN or an
$e^+e^-$ Linear Collider (*LC) to explore the remainder of the parameter
space.

\subsection{Searches at Hadron Supercolliders}

At the LHC, the production of gluinos and squarks remains the dominant
mechanism for sparticle production for masses in excess of 1~TeV. The
cascade decays of the gluinos and squarks lead to multi-jet plus
multi-lepton plus $\eslt$ event topologies.
In the $\eslt$ channel, the LHC can search\cite{ATLAS,BCPT2}
for gluinos and squarks
with
masses between 300~GeV up to about 1.3~TeV (2.0~TeV)
for $m_{\tq}>>m_{\tg}$ ($m_{\tq}=m_{\tg}$). It is instructive to note that
the same sign dilepton and
several multilepton signals {\it must} simultaneously be present\cite{BTW}
if any
$\eslt$ signal is to be attributed to squark and gluino production, though
their
relative rates could be sensitive to the entire sparticle spectrum.
The rate for like-sign dilepton\cite{BTW,BGH} plus $\eslt$ events is enormous
for $m_{\tg} \leq 300$~GeV; this ensures there is no window between the
Tevatron and the LHC where gluino of the MSSM may escape
detection\cite{BTW,ATLAS}.
Gluinos and squarks with masses up to about 1~TeV should be detectable in
these channels.
Gluinos
and squarks may also be a source of high $p_T$ $Z + \eslt$ events at the
LHC.

\begin{table}
\small
\caption[] {Estimates of the discovery
reach of various options of future hadron colliders. The signals have
mainly been computed for negative values of $\mu$. We expect that the reach
in especially the $all \to 3\ell$ channel will be sensitive to the sign of
$\mu$. The two entries for the $\eslt$ channel refer to results of two
independent analyses.}

\vskip 0.5\baselineskip
\footnotesize
\begin{tabular}{c|ccccc}
&Tevatron II&Main Injector&Tevatron$^*$&DiTevatron&LHC\\
Signal&0.1~fb$^{-1}$&1~fb$^{-1}$&10~fb$^{-1}$&
1~fb$^{-1}$&10~fb$^{-1}$\\
&1.8~TeV&2~TeV&2~TeV&4~TeV&14~TeV\\
\hline
$\eslt (\tq \gg \tg)$ & $\tg(210)/\tg(185)$ &
$\tg(270)/\tg(200)$ & $\tg(340)/\tg(200)$ & $\tg(450)/\tg(300)$
& $\tg(1300)$ \\

$l^\pm l^\pm (\tq \gg \tg)$ & $\tg(160)$ & $\tg(210)$ &
$\tg(270)$ & $\tg(320)$ & \\

$all \rightarrow 3l$ $(\tq \gg \tg)$ & $\tg(180)$ &
$\tg(260)$ & $\tg(430)$ & $\tg(320)$ & \\

$\eslt (\tq \sim \tg)$ & $\tg(300)/\tg(245)$ &
$\tg(350)/\tg(265)$ & $\tg(400)/\tg(265)$ & $\tg(580)/\tg(470)$ &
$\tg(2000)$ \\

$l^\pm l^\pm (\tq \sim \tg)$ & $\tg(180-230)$ & $\tg(320-325)$ &
$\tg(385-405)$ & $\tg(460)$ & $\tg(1000)$ \\

$all \rightarrow 3l$ $(\tq \sim \tg)$ & $\tg(240-290)$ &
$\tg(425-440)$ & $\tg(550^*)$ & $\tg(550^*)$ &
$\stackrel{>}{\sim}\tg(1000)$ \\

$\tilde{t}_1 \rightarrow c \tz_1$ & $\tilde{t}_1(80$--$100)$ &
$\tilde{t}_1 (120)$ & \\

$\tilde{t}_1 \rightarrow b \tw_1$ & $\tilde{t}_1(80$--$100)$ &
$\tilde{t}_1 (120)$ & \\

$\Theta(\tilde{t}_1 \tilde{t}_1^*)\rightarrow \gamma\gamma$ &
--- & --- & --- & --- & $\tilde{t}_1 (250)$\\

$\tl \tl^*$ & $\tl(50)$ & $\tl(50)$ & $\tl(50)$ & &
$\tl(250$--$300)$

\end{tabular}
\end{table}

Experiments at the LHC can also search for ``hadron-free'' trilepton
events from
$\tw_1\tz_2$ production. Backgrounds from top quark production can be very
effectively suppressed\cite{LHCWIN} by requiring that the two hardest
leptons have the
same sign of charge. The signal again becomes unobservable when the two-body
spoiler modes $\tz_2\to (Z$ or $H_{\ell}) + \tz_1$ become accessible.
For positive values of $\mu$ there still remain holes in the parameter space
where this signal is unobservable, even if the spoiler channels are closed.
The dilepton
mass distribution in $\ell^+\ell^-\ell'$ events can be used to reliably
measure $m_{\tz_2}-m_{\tz_1}$.

Selectrons and smuons with masses up to 250~GeV (300~GeV
if it is possible to veto central jets with an efficiency of 99\%) should
also be detectable\cite{SLEP}. Finally, we note that
with an integrated luminosity
of 100 $fb^{-1}$ the $\gamma\gamma$ decays of scalar stoponium has been
argued to allow for the detection of $\tt_1$ with a mass up to 250~GeV,
assuming $\tt_1\to b\tw_1$ (and, perhaps, also $\tt_1 \to bW\tz_1$)
is kinematically forbidden\cite{DN}.

The reach of various proposed and approved hadron collider facilities is
summarized in Table 1. Although we have not discussed it explicitly, we have
also included the DiTevatron, a proposed 4~TeV upgrade of the Tevatron, and
assumed an integrated luminosity of 1 $fb^{-1}$.
This Table\cite{DPF} clearly reiterates our earlier point that weak scale
supersymmetry can only be thoroughly explored at supercolliders.

\subsection{$e^+e^-$ Supercolliders}

Charged sparticles (and sneutrinos, if $m_{\tnu} > m_{\tw_1}$)
should be readily detectable at the *LC if their pair production is
kinematically accessible\cite{JLC}. Unfortunately, most detailed $e^+e^-$
studies to date do not incorporate the cascade decays\cite{BBKMT}
of sparticles.\footnote{The production and cascade decays of sparticles
by $e^+e^-$ collisions has now been incorporated into ISAJET, as well as
in SUSYGEN, a Monte Carlo program\cite{KATSEN} that includes initial state
radiation and interfaces with the LUND JETSET hadronization routines. Both
ISAJET and SUSYGEN currently lack spin correlations, final state decay
matrix elements and the possibility of simulations with polarized beams.}
These machines are also an ideal facility for detection and a study
of the properties of the Higgs sector. While the detection of Higgs bosons
beyond the reach of LEP II is possible at the LHC (except possibly for the
well-known ``hole'' in the $m_{H_p}-\tan\beta$ parameter plane) it requires
excellent performance of both the machine and the detectors\cite{ATLAS,CMS}.
In contrast, the detection of Higgs bosons at these facilities is relatively
easy, and
for favourable values of parameters, it may even be possible to distinguish
the Higgs sector of the SM and that of SUSY\cite{JANOT}.

A 500~GeV $e^+e^-$ collider will be able to search for charginos
essentially all the way up to the beam energy. If we assume gaugino
mass unification, this corresponds to a gluino mass reach of about
700-800~GeV. The real
power of these machines, however, lies in the ability to
do precision experiments which can then be used to probe\cite{MUR} unified
models of interactions as discussed shortly.

\section{Supergravity Phenomenology}

We have seen that SUGRA GUT models provide an attractive and economic
framework for SUSY phenomenology. Their economy derives from the fact
that with well-motivated assumptions about physics at an
ultra-high scale, all sparticle properties are determined by just the
small parameter set (1).
It is remarkable that even the simplest SUGRA GUTs are
consistent with experimental constraints as well as
cosmology\cite{LOPEZ,BDKNT}. Nevertheless, since the model is based on
untested assumptions, it is essential to devise tests that can,
in principle, falsify the framework.
\begin{figure}[htb]
\vspace{170mm}
\caption{Regions of the supergravity parameter space that can be explored at
the Tevatron and its Main Injector upgrade or at LEP II. The region below
the hatched line is already excluded by current experiments. The 3$\ell$
reach, in this figure, is only approximate since it was not obtained from
a complete simulation. This may be compared to the corresponding reach
shown in Fig.~2 obtained from a detailed simulation of trilepton events.
The region that can be probed via $\eslt$ and multilepton signals
from gluinos and squarks (see Table 1) may roughly be obtained from the
right hand scale in the figure which has been taken from Ref.[14].}
\end{figure}

As mentioned in Sec. 1, since the phenomenology is determined in terms of just
four SUSY parameters, various SUSY
cross sections become
correlated. Different experimental analyses
from $e^+e^-$ and hadron
colliders
can thus be
consistently combined into a single framework\cite{BCMPT}; since various
searches frequently probe different parts of the parameter space
they often complement one another. This is illustrated in the $m_0-m_{1/2}$
plane in Fig.~3 for the
Tevatron\cite{BCMPT}, and in Fig.~4 for the LHC\cite{BCPT2}. The grey region
(brick-covered region)
in Fig.~3 (Fig.~4) is excluded by theoretical constraints: either the
correct electroweak symmetry breaking pattern with the experimental value of
$M_Z$ is not obtained, or the lightest neutralino is not the LSP. The hatched
region is excluded by current experiments. The dashed and dashed-dotted lines
in Fig.~3 denote the regions where sleptons and charginos might be detectable
at LEP II or the Tevatron, while the dotted lines mark the boundaries of
the ``spoiler modes'' of the $\tz_2$. The corresponding regions along
with the boundary of the region where sleptons might be detectable at the
LHC (denoted as $\tell_R(200))$ as well as the boundary of the region where
the $\eslt$ signal from gluinos and squarks is observable above background
at the $5\sigma$
level is shown in Fig.~4. We see that before the advent of the
supercollider era, the various signals in Fig.~3 probe complementary
regions of the parameter plane (the reach in the $\eslt$ channel is
200-250~GeV depending on the squark mass). In contrast, Fig.~4
shows that the $\eslt$ signal probes much farther than the other signals
at the LHC. It should be kept in mind, that part of this plane may also
be probed via the same sign and multilepton signals from gluinos and squarks.
This is under study. We note here that enhanced decays of gluinos
to third generation fermions and sfermions discussed in Sec. 1 are an
important source of multilepton events, over and above the usually studied
gluino and squark cascade decays to charginos and neutralinos.
\begin{figure}[htb]
\vspace{170mm}
\caption{The regions of the $m_0-m_{1/2}$ plane that might be explored in
various channels at the LHC with an integrated luminosity of 10~$fb^{-1}$.
The clean $3\ell$ signal allows the exploration of most of the plane
below the boundary of the spoiler mode if $\mu < 0$. For positive values of
$\mu$, the situation is similar to Fig.~2 in that there are holes for
400~GeV $\alt m_0 \alt 1000$~GeV. Also shown are regions where slepton signals
should be observable at the LHC and where chargino and Higgs boson signals
might be observable at LEP II. This figure is from Ref.[37].}
\end{figure}

At $e^+e^-$ supercolliders, it has been shown that it should be
possible\cite{JLC,MUR}
to measure the masses of charginos, sleptons
with a precision of a few GeV, assuming
that they directly decay to the LSP.
It has also been shown that a similar
precision can be attained in the measurement of squark masses\cite{FF}.
These mass measurements should make it
possible to test the assumed unification of scalar masses in SUGRA models:
Unmixed squark and sleptons masses should be generation
independent.\footnote{This is important since an alternative
proposal\cite{SEIB} for
suppressing flavour changing neutral currents does not require squarks to be
degenerate.} They
are related to one another if we further assume the desert hypothesis as
in minimal SUGRA GUTs, but to test this relation, knowledge of gaugino masses
is necessary.

The availability of polarized beams essentially doubles the number of
observables and allows for other interesting tests of this framework. For
example, if $\te_R$ and $\tw_1$ are both kinematically accessible, a
global fit from the measurement of $m_{\tw_1}$, $m_{\tz_1}$, $m_{\te_R}$,
together with cross sections
$\sigma_R(\te_R\te_R)$ and $\sigma_R(\tw_1\tw_1)$ from right-handed electron
beams,
would allow the
determination of the two electroweak gaugino masses, $\tan\beta$ and $\mu$
(with a precision depending on where we are in parameter space). This enables
a test of the unification of gaugino masses at about the 5\% level\cite{MUR}.

We thus see that the observation
of sparticles would not only be a spectacular new discovery, but a measurement
of their properties, particularly at future colliders would test various
SUGRA assumptions, and so, serve as a telescope to the unification scale.

\section{More Ambitious Measurements}

\subsection {How do we tell it is SUSY?}
It is well known that the clean environment of $e^+e^-$ colliders
allows for precision studies of any new physics that might be accessible.
We have already seen that it should be possible to measure sparticle masses
to a precision of a few percent. Measurements of angular distributions readily
allow for the determination of particle spin\cite{DION,JLC}, at least for
the lighter sparticles that directly decay to the LSP. The observation of
several new particles with the spin, charge and colour quantum numbers of
the predicted sparticles would be very strong evidence for supersymmetry.
However, the most definitive test of (softly broken)
supersymmetry comes from the fact that, exactly as
in spontaneously broken gauge theories, the relationships between the
dimensionless couplings implied by the (super)symmetry
continue to be preserved, while the corresponding relationships between
sparticle masses might be badly violated. For instance, the coupling of
the photino to the lepton-slepton system is, apart from a Clebsch-Gordan
coefficient, exactly the same as the usual electromagnetic coupling of the
photon to an electron pair.
Such relationships between various couplings in a softly broken SUSY
theory are direct consequences of the underlying supersymmetry, and differ
from the tests discussed in the last section. These latter assume
supersymmetry and test the various assumptions about the symmetries of
interactions at the high scale that form the underpinnings of SUGRA GUT models,
but do not directly test SUSY.

These direct tests of SUSY
are complicated by the fact that generally speaking, the
physical particles are unknown mixtures of states with different
SU(2)$\times$U(1) quantum numbers. It is only after the mixing patterns are
untangled can the SUSY relations be tested. Obviously, this requires precise
measurements of a large number of observables.

The feasibility of testing SUSY relations between the couplings
at a
500~GeV $e^+e^-$ collider
has been examined by Feng {\it et. al.}\cite{FENG}.
Here, we illustrate the basic idea with an example involving pair production
of charginos.
The starting point is that
assuming only the MSSM field content but not supersymmetry, the
$2\times 2$ chargino
mass matrix can obviously be written in terms of four real parameters. They
then observe that for
the case where the chargino is a substantial mixture of the Higgsino
and the gaugino,
both $\tw_1\tw_1$ and $\tw_2\tw_1$ production is kinematically
possible at a 500~GeV
linear collider if the masses are as in the MSSM. Based on earlier
studies\cite{JLC,MUR} they assume that these chargino masses will be
accurately measured.
Next, they observe that a measurement of $\sigma_R(\tw_1\tw_1)$,
the pair production cross section for right-handed electron beam (this
does not couple to sneutrinos), and a suitably defined\cite{FENG}
forward-backward asymmetry suffices to determine the gaugino-Higgsino
mixing angles $\phi_L$ and $\phi_R$ of the two chiral components of $\tw_1^-$.
The experimental determination of the two masses and the two mixing angles
is then used to determine the four parameters of the chargino mass
matrix\footnote{Further details as well as other
complications not discussed here are may be found in the original
paper\cite{FENG} along
with other tests of supersymmetry}.
As is well known, the squared sum of the off diagonal elements is fixed
by supersymmetry to be $2M_W^2$, so that an experimental verification
of this would indeed amount to a direct confirmation of supersymmetry.
It is shown that with an integrated
luminosity of 100 $fb^{-1}$ it should be possible to test
this relation with a precision of 15-20\%. A different set of
measurements\cite{FENG} serves to test the SUSY relation with similar
precision in the case that the chargino is dominantly a gaugino.

\subsection{Mass Measurements at Hadron Supercolliders}
As we have already mentioned, it is possible to make rather precise
measurements of sparticle masses in the clean environment of an $e^+e^-$
collider. However, unlike the LHC, the construction of these machines has
not yet been approved, and it is possible that these may become operational
several years after the LHC commences operation.\footnote{In this case, it
would be important to archive the LHC data in appropriate format for
subsequent re-analysis, particularly of cascade decays,
in light of new information on chargino and neutralino properties
from experiments at the $e^+e^-$ collider.} It is then of interest to ask
whether it is possible to obtain some measure of sparticle masses despite
the relatively messy environment of hadron collisions.
This would be possible if one can isolate a relatively clean
sample of signal events from a {\it single} SUSY process, {\it i.e.}
relatively free both from SM backgrounds {\it and} with
only small contributions from other SUSY processes. An analysis of event shape
variables would then lead to some information about SUSY masses.

One process for which this strategy has been shown\cite{LHCWIN} to work rather
well is for the clean trilepton signal from $\tw_1\tz_2$ production. These
authors have identified a set of cuts which not only eliminate the SM
backgrounds, but also reduce the contributions from gluino and squark
cascade decays as well as from associated production processes
(which would otherwise dominate the trilepton signal) to acceptable
levels. For the case of negative $\mu$ which they studied in detail,
sufficiently many signal events remain after the cuts so that the end point
of the $m(\ell\bar{\ell})$ distribution in $\ell\bar{\ell}\ell'$ events
yields a reliable measure of $m_{\tz_2}-m_{\tz_1}$. A study of other
distributions may provide constraints on other combinations of the chargino
and neutralino masses.

Is it possible to measure the mass of the gluino which will perhaps be the
most copiously produced sparticle? Several strategies have been proposed in the
literature.
In the first instance\cite{ASSOC}, associated production processes such as
$\tg\tz_1$ were examined using parton level event generators. If these
events could be singled out, then the ambiguities from producing and
decaying {\it two} strongly interacting sparticles are by-passed. In
practice, rather hard cuts were required to separate the $\tg\tz_1$
events from $\tg\tg$ events.  Ultimately, it was concluded that this
reaction might be of use only if $m_{\tg}\alt 350$ GeV.
In a subsequent study\cite{BGH}, it was claimed that $m_{\tg}$ could be
measured to $15\%$ by focussing on same-sign isolated dilepton events
from $\tg\tg$ production, where each gluino decays via $\tg\to
q\bar{q}\tw_1$, with $\tw_1\to \ell\nu_{\ell}\tz_1$. The idea was to divide
each event into two hemispheres and use the larger of the invariant masses
in the two hemispheres as an estimator of $m_{\tg}$.
However, these
conclusions may be overly optimistic, since this study considered
only a single production mechanism ($\tg\tg$ production-- whereas same
sign dileptons can come from various SUSY sources), a single cascade
decay chain, and neglected effects of additional QCD radiation.
Very recently, it has been shown\cite{BCPT2}, that it may be possible to
measure $m_{\tg}$ from the $\eslt$ sample by using eigenvectors for the
construction of the transverse sphericity to again\cite{BGH} divide the event
into two halves
and constructing the invariant mass of the jets in each hemisphere.
With carefully chosen cuts, the
distributions of $M_{est}$, the larger of these two invariant masses,
are sufficiently distinct for values of $m_{\tg}$ differing by 15-25\%,
depending on $m_{\tq}$.
Monte Carlo studies should, therefore,
enable a measurement of $m_{\tg}$ with roughly
the same precision. This last analysis includes the production of all
sparticles
in the simulation as well as contamination to the jet masses from QCD
radiation. It would be interesting to test if the method survives a complete
detector simulation.

Finally, one may ask whether it is possible to tell what sparticles are
contributing to a SUSY event sample. For the case of clean trileptons where
it is possible to devise cuts to minimise other SUSY sources as just
discussed, this is moot. A more interesting question is to ask if the
production of squarks makes a significant sample to the $\eslt$ and
multilepton samples from SUSY. Possible strategies for identifying the
presence of squarks
in SUSY events include:

\begin{itemize}

\item A study of the charged asymmetry in the same sign dilepton sample ---
more up type squarks are expected to be produced in $pp$ collisions, so
that the number of $\ell^+\ell^+$ events should be larger than $\ell^-\ell^-$
events\cite{BTW}. This might be possible
if gluinos and squarks are indeed the main
source of same sign dileptons, but this is not guaranteed.

\item Since gluinos decay via three body processes and frequently
have lengthier cascades than squarks ($\tq_R$ often decays directly to the
LSP), it is reasonable to expect a higher jet multiplicity for
the case when $m_{\tq} >> m_{\tg}$, {\it i.e.} when $\tg\tg$ production
dominates. This has been studied\cite{BCPT2} using ISAJET and it has been found
that
despite contamination from QCD radiation, the $\eslt$ sample generated
with $m_{\tq} >> m_{\tg}$,
exhibits about half a unit larger multiplicity than the corresponding
sample with $m_{\tq}\sim m_{\tg}$. Of course, the jet multiplicity
distributions are sensitive to the sparticle masses, so that it is
necessary to have some idea of these masses before conclusions can be drawn.

\end{itemize}

There are many other questions one may ask. Can one tell the presence of
third generation squarks in SUSY events? Can one tell the presence of
Higgs bosons produced via SUSY cascades? The answers to both
questions are positive for at least some regions of parameter space. The
multiplicity of tagged $b$-jets\cite{BTW,BARTL} provides a good handle for
this purpose. In fact, in some regions, the $H\to b\bar{b}$ signal may be
identifiable\cite{BCPT2} as a bump in the invariant mass ditribution of tagged
$b$ jet
pairs. For this, one obviously needs a large number of tagged $b$ jets from
Higgs boson decay in events that do not simultaneously contain other
sources of $b$-jets which would produce a large combinatorial background.
Another requirement for the observation of the Higgs bump is that the rate
from other sources of $b$-jets ({\it e.g.} from $\tg \to \tt_1 t$ or $\tb_1 b$
events, where the squarks may be real or virtual) should not be overwhelmingly
large.

Hadron colliders are generally regarded as discovery machines, and it is
widely believed that it will be very difficult
to measure the masses or elucidate the physics of any new particles that are
produced. However, the examples given above show that this is not
entirely correct. Such studies have only just begun, and it remains to be seen
just what other information can be gleaned from new physics event samples
in spite of the messy environment
of hadron colliders.

\section{Final Remarks}
To conclude, if low energy supersymmetry is indeed the physics that
stabilizes the electroweak scale, it will be unlikely to escape detection at
the LHC or at future $e^+e^-$ colliders with $\sqrt{s} \stackrel{>}{\sim}
500-800$~GeV. While Tevatron upgrades indeed probe substantial regions of SUSY
parameter space, a non-observation of any signal will not exclude weak scale
SUSY. Supercolliders appear to be essential for a comprehensive exploration
of the parameter space as well as for the elucidation of any signals that
may be found. We cannot overstress the complementary nature of $e^+e^-$ and
hadron supercolliders if SUSY is present at the weak scale. Experiments at
these facilities not only will unambiguously confirm or exclude the existence
of
sparticles, but will allow a comprehensive study of their properties which,
in turn, may provide us information about physics at an
ultra-high energy scale.

\vspace{4mm}
\noindent {\bf Acknowledgements:}
It is a pleasure to thank Asim Ray, Ranabir Dutt and their colleagues for
arranging this meeting in the marvellous atmosphere of Santiniketan,
and especially for their unrelenting
efforts to accommodate the needs of various
participants.
Collaborations and dicussions
with H.~Baer, A.~Bartl, M.~Brhlik, M.~Bisset, C-H.~Chen,
M.~Drees, R.~Godbole, J.~Gunion, C.~Kao, R.~Munroe, H.~Murayama,
M.~Nojiri, F.~Paige,
S.~Protopopescu, J.~Sender and J.~Woodside from whom I have learnt much
are gratefully acknowledged. This
research was supported by the U.S. Dept. of Energy grant DE-FG-03-94ER40833.

\bibliographystyle{unsrt}

\end{document}